\documentclass[preprint,twocolumn,showpacs,superscriptaddress]{revtex4-1}%
\usepackage{amsmath}
\usepackage{amssymb}
\usepackage{bm}
\usepackage{epsfig}
\usepackage{graphicx}
\usepackage{color}
\usepackage[colorlinks=true]{hyperref}
\usepackage[T2A]{fontenc}
\usepackage[cp1251]{inputenc}
\usepackage[russian,english]{babel}
\usepackage{amsfonts}%
\providecommand{\U}[1]{\protect\rule{.1in}{.1in}}
\providecommand{\U}[1]{\protect\rule{.1in}{.1in}}

\graphicspath{{Images/}}
\begin{document}
\title{Dynamical Generation of Spin Current and Phase Slip in Exciton-Polariton Condensates}
\author{$^{\ast}$Bo Xiong}
\affiliation{Department of Physics, Nanchang University, 330031 Nanchang, China}
\affiliation{Skolkovo Institute of Science and Technology, Novaya Street 100, Skolkovo
143025, Russian Federation}
\email{stevenxiongbo@gmail.com}
\email{ORCID iD: 0000-0003-2434-4898}

\begin{abstract}
We show that how to generate propagation of spin degree in spin-symmetric
exciton-polariton condensates in a semiconductor microcavity. Due to the
stimulated spin-dependent scattering between hot excitons and condensates,
exciton polaritons form a circular polarized condensate with spontaneous
breaking of the spin rotation symmetry. The spin antiferromagnetic state is
developed evidently from the density and spin flow pumped by localized laser
source. The low energy spin current is identified where the steady state is
characterized by the oscillating spin pattern. Finally, we predict via
simulation how to dynamical generation of phase slip where ring-shape phase
jump shows the behavior of splitting and joining together.

\end{abstract}
\date{\today }

\maketitle

\section{Introduction.}

Recently, in semiconductor microcavities with quantum wells sandwiched between
highly reflective mirrors, the strong coupling is achieved between excitons
and photons~\cite{WeisbuchPRL1992,microcavities,DengRMP2010,CarusottoRMP2013}.
Such coherent light-matter particles called exciton-polaritons obey the
Bose-Einstein statistics and thus condense at critical temperatures ranging
from tens Kelvin~\cite{DangNature2006,DengScience2002,BaliliScience2007} till
several hundreds Kelvin~\cite{ChristmannAPL2008, XiePRL2012}, which exceeds by
many orders of magnitude the Bose-Einstein condensation temperature in atomic
gases. Recently, electrically pumped polariton laser or condensation was
realized based on a microcavity containing multiple quantum wells
\cite{WertzNphys2010,SchneiderNature2013}. Considering the high transition
temperatures and high tunability from pumping source, semiconductor
microcavities are perfectly suited for studies of macroscopically collective
phenomenon and have initiated the fascinating research on the polariton
quantum hydrodynamics.

The polaritons have two allowed spin projections on the structure growth axis,
$\pm$1, corresponding to right- and left- circular polarizations of photons.
In diverse semiconductor materials like GaAs/GaAlAs \cite{LouNP2007}, Si
\cite{DashNature2009}, organic single-crystal microcavity SiN$_{x}$/SiO$_{2}$
\cite{Kera-CohenNP2010} and so on, spin injection and detection has been
successfully realized which is one of the key ingredients for functional
spintronics devices. A number of prominent spin-related phenomena both in
interacting and in noninteracting polariton systems have already been
predicted and observed in the microcavities, such as, spontaneous polarization
\cite{SnokeScience2002,BaumbergPRL2008,BaumbergPRX2015,BaumbergPRL2016,AskitopoulosPRB2016,sigurdsson2017driven,krol2018spin}%
, polarization
multistability~\cite{GippiusPRL2007,ParaisoNM2010,Ouellet-PlamondonPRB2016,ohadi2017spin,kulakovskii2018elliptically,klaas2019nonresonant}%
, optical spin Hall
effect~\cite{KavokinPRL2005,LeyderNphys2007,KammannPRL2012,FlayacPRl2013,NalitovPRL2015,DufferwielPRL2015,SalaPRX2015}
and topological insulator
\cite{NalitovPRL2015B,BleuPRB2016,gulevich2016kagome,kartashov2017bistable,klembt2018exciton,kozin2018topological}%
, spin Zeeman and Meissner effect\textbf{ }%
\cite{rousset2017magnetic,mirek2017angular,krol2019giant}.

Spin degrees of freedom in two-dimensional exciton-polaritons superfluid can
drastically change elementary topological vortices referred to as half-quantum
vortices (HQV)
\cite{Volovik2003,RuboPRL2007,LagoudakisScience2009,SanvittoNphys2010,HivetNphys2012}
which are characterized by a half-integer value of vorticity in contrast to
the regular quantum vortex
\cite{LagoudakisNphys2008,RoumposNphys2011,NardinNphys2011,SanvittoNphot2011,DominiciSA2015,BoulierPRL2016,caputo2016topological,caputo2019josephson}
where the vorticity takes only integer values. Usually HQV carry only one
half-integer topological charge originating both from the superfluid current
proportional to $\nabla\theta$, and from $\pi$ spin disclinations superimposed
as a result of Berry's phases induced by spin rotations \cite{Thouless1998}.
Relevant ideals of half vortices have been discussed in A phase of $^{3}$He
\cite{LeggettRMP1975,SalomaarRMP1987,Vollhardt1990}, in triplet
superconductors Sr$_{2}$RuO$_{4}$ \cite{MackenzieRMP2003} and spinor atomic
Bose-Einstein condensates
\cite{HoPRL1998,MachidaJPSJ1998,ZhouPRL2001,MukerjeePRL2006} with two
different spin components where HQV is just residing in one of components
\cite{KirtleyPRL1996,YamashitaPRL2008,JangScience2011,ChoiPRL2012,SeoPRL2015}.

However, precise coherent control of spin polarization, propagation and
topological defects in exciton-polariton condensates still remains a core
challenge. Here, we address this problem, and demonstrate exciton-polariton
condensates will not only show spontaneous polarization and also coherent
propagation of the pseudospin under nonlocal spin injection. When taking into
account incoherent hot exciton reservoir scattered into coherent states,
dramatically enhanced spin-polarized signal can be observed at the appropriate
pumping regime. Moreover, the coherent spin antiferromagnetic state can also
be identified and manipulated by spin-symmetric pumping source. Additionally,
cavity engineering allows us to the dynamic generation of phase slip where
ring-shape phase jump shows the behavior of splitting and joining together
induced by incoherent reservoir as a result of effective gauge field.

\section{Physical Background.}

In the absence of external magnetic field the \textquotedblleft
spin-up\textquotedblright\ and \textquotedblleft spin-down\textquotedblright%
\ states $\sigma=\pm$ of noninteracting polaritons, or their linearly
polarized superpositions, are degenerate corresponding to the right $\left(
\sigma_{+}\right)  $ and left $\left(  \sigma_{-}\right)  $ circular
polarizations of external photons. The spinor nature of exciton polaritons can
therefore be manifested since the spin are essentially free in semiconductor
microcavities. To illustrate the fully degenerate spinor nature, and as a
first step, the Zeeman energy must be much smaller than the interaction
energy. Thus we shall consider only the case of zero magnetic field achieving
a good approximation in the following. Since the interaction between exciton
polaritons depends on their total spins (singlet or triplet), their spin
states may be changed after the scattering. The spin-dependent interactions
cause the polariton spin states exchange. Moreover, additional mixing may
comes from the longitudinal-transverse (LT) splitting of polaritons (referred
to as the Maialle mechanism) \cite{PanzariniPRB1999} and from structural
anisotropies \cite{DasbachPRB2005}.

The low energy dynamics is therefore described by a pairwise interaction that
is spin-rotation invariant and preserves the spin of the individual exciton
polaritons. The general form of this interaction is $\hat{V}\left(
\mathbf{r}_{1}-\mathbf{r}_{2}\right)  =\delta\left(  \mathbf{r}_{1}%
-\mathbf{r}_{2}\right)
%TCIMACRO{\dsum \nolimits_{F=0}^{2f}}%
%BeginExpansion
{\displaystyle\sum\nolimits_{F=0}^{2f}}
%EndExpansion
g_{F}\cdot\hat{P}_{F}$ where $g_{F}=4\pi\hbar^{2}a_{F}/M$, $M$ is the mass of
exciton polaritons, $\hat{P}_{F}$ is the projection operator which projects
the pair 1 and 2 into a total spin F state, and $a_{F}$ is the s-wave
scattering length in the total spin F channel. For exciton polaritons of $f=1$
bosons, interaction has form $\hat{V}=g_{0}\cdot\hat{P}_{0}+g_{2}\cdot\hat
{P}_{2}$. In terms of nonlinear optics, the coupling coefficients of
polarization independent $c_{0}$ and so-called linear-circular dichroism
$c_{2}$ can be estimated through the matrix elements of the
polariton-polariton scattering in the singlet and triplet configurations.

It is convenient to write the Bose condensate $\Psi_{a}(\mathbf{r})\equiv
<\hat{\psi}_{a}(\mathbf{r})>$ as $\Psi_{a}(\mathbf{r})=\sqrt{n(\mathbf{r}%
)}\zeta_{a}(\mathbf{r})$, where $n(\mathbf{r})$ is the density, and $\zeta
_{a}$ is a normalized spinor $\zeta^{+}\cdot\zeta=1$. It is obvious that all
spinors related to each other by gauge transformation $e^{i\theta}$ and spin
rotations $\mathcal{U}(\alpha,\beta,\gamma)=$$e^{-iS_{x}\alpha}e^{-iS_{y}%
\beta}e^{-iS_{z}\gamma}$ are degenerate, where $(\alpha,\beta,\gamma)$ are the
Euler angles.

The non-equilibrium dynamics of polariton condensates is described by a
Gross-Pitaevskii type equation for the coherent polariton field, which should
be coupled to a hot-excitons reservoir excited by the nonresonant exciting
pump. The model is, however, generalized to take into account the polarization
degree of freedom of hot exciton. In this approach, instead of polarization
independent scattering, we must take into account dichroism scattering between
hot exciton and coherent polariton field.

Let us turn to the pseudospin representation, then the local spin density
$\overrightarrow{s}$ at the position $\mathbf{r}$ and time $t$ is
$\overrightarrow{s}\left(  \mathbf{r},t\right)  =\Psi^{\dag}(\mathbf{r,t}%
)\hat{\overrightarrow{s}}\Psi(\mathbf{r,t})$, where $\hat{\overrightarrow{s}%
}=\left(  \hbar/2\right)  \hat{\overrightarrow{\sigma}}$ with $\hat
{\overrightarrow{\sigma}}$ being the Pauli matrices. The usual definition of
the free-particle probability current $\mathbf{J}_{n}=\operatorname{Re}\left[
\Psi^{\dag}(\mathbf{r,t})\frac{\mathbf{\hat{P}}\hat{I}}{m}\Psi(\mathbf{r,t}%
)\right]  $, where $\hat{I}$ is the identity, and probability spin current
$\mathbf{J}_{\overrightarrow{s}}=\operatorname{Re}\left[  \Psi^{\dag
}(\mathbf{r,t})\frac{\mathbf{\hat{P}}\overrightarrow{s}}{m}\Psi(\mathbf{r,t}%
)\right]  $. In addition, the emergent magnetic monopoles defined by analogy
with Maxwell's equation as $\nabla\cdot\overrightarrow{s}$ can be realized and
characterized by a divergent in-plane pseudospin pattern, that have been
present in magnetically frustrated materials, spin-ice
\cite{HarrisPRL1997,CastelnovoNature2007,JaubertNatPhys2009,FennelScience2009,MorrisScience2009,KadowakiJPSJ2009,BramwellNature2009}%
, magnetic nanowires \cite{ParkinScience2008} and atomic spinor Bose-Einstein
condensates \cite{BuschPRA1999,StoofPRL2001}. The dynamics of each spin under
the effect of magnetic field is governed by the precession equation
$\partial_{t}{\mathbf{S}}={\mathbf{H}}\times{\mathbf{S}}/\hbar$. The total
effective magnetic field $\mathbf{H}$ represents the sum of the field
responsible provided by the spin dependent and independent polariton-polariton
interactions and polariton-hot exciton interaction (LT splitting
$\mathbf{H}_{LT}$ is assumed to be negligible in high density regime). Very
different from those isolated or closed system, the dynamic of spin pattern in
such open-dissipative system is crucially determined by the pump source. We
will go into further details in the following.

\begin{figure}
[t]%
\includegraphics[width=0.9\linewidth]{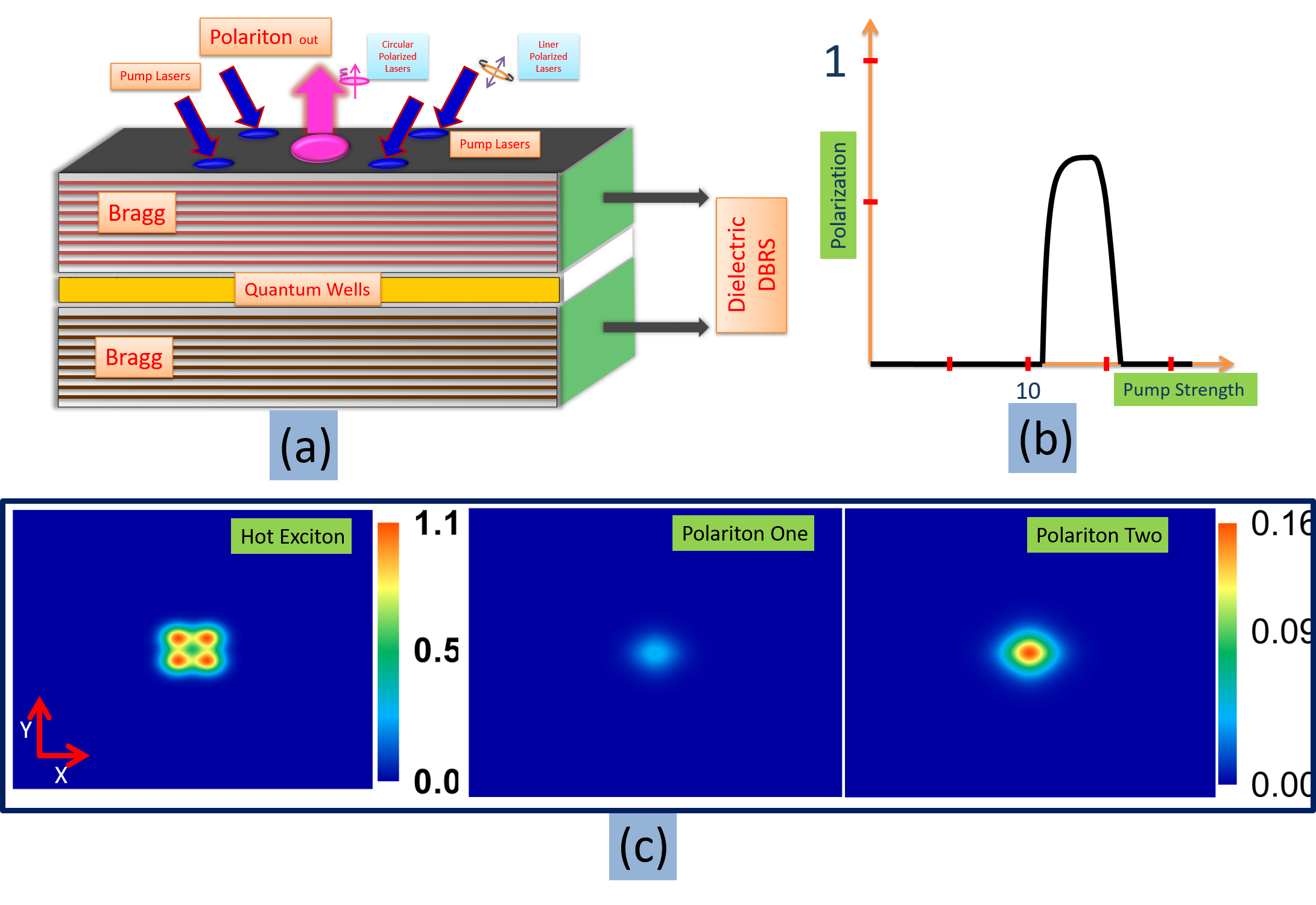}\caption{ (Color online) The
spontaneously circular polarization of spinor condensate non-resonant pumped
by linearly polarized laser. (a) Proposed scheme to experimentally stimulating
spontaneous circular polarization by nonpolarized laser beam. (b) Spinor is
polarized when the laser power is larger than first threshold value, however,
unpolarized after laser power is above second threshold value. (c) Density
distribution of hot exciton (left picture which has the same profile for both
components) and spinor polariton (middle and right pictures for each
components) in real space. Here, simulations are in the absence of disorder
for 4 pumping points with a small radius $1.54$ $\mu$m. The size of profile is 24x24
and the other parameters used in the simulations are shown in the paper.}\label{Fig0}%

\end{figure}

\section{Theoretical Model.}

In the following, we study the propagation of polarized polariton in the a
planar microcavity and generation of spin polarization, spin current and the
observability of the HQV, in realistic structures. The equation of motion for
the spinor polariton wave function reads
\cite{CarusottoPRL2004,LittlewoodPRL2006,WoutersPRL2007,WoutersPRB2008}
\begin{widetext}%
\begin{align}
i\hbar\partial_{t}\psi_{\pm}\left(  \mathbf{r}\right)   &  =\left\{
-\frac{\hbar^{2}}{2m}\nabla^{2}+\frac{i\hbar}{2}\left(  g_{2}n_{R\pm}%
+h_{2}n_{R\mp}+\beta_{2}|\psi_{\pm}|^{2}+f_{2}|\psi_{\mp}|^{2}-\gamma
_{C}\right)  +V_{ext}\left(  \mathbf{r}\right)  \right\}  \psi_{\pm}\left(
\mathbf{r}\right) \nonumber\\
&  +\left\{  \hbar\left(  \beta_{1}|\psi_{\pm}|^{2}+f_{1}|\psi_{\mp}%
|^{2}\right)  +V_{R}\left(  \mathbf{r}\right)  \right\}  \psi_{\pm}\left(
\mathbf{r}\right)  , \label{Pola_1}%
\end{align}
\end{widetext}where $\psi_{\sigma}$ represents the condensed field, with
$\sigma=\pm$ representing the spin state of polaritons with effective mass
$m$. $\gamma_{C}$ represents the coherent polariton decay rate. $\beta_{1}$
and $f_{1}$ is the spin-conserved and spin-exchange polariton-polariton
interaction strength, respectively. $n_{R\sigma}$ is the density of the
incoherent hot exciton reservoir. And here, $V_{R}\left(  \mathbf{r}\right)
=\hbar\left[  g_{1}n_{R\pm}+h_{1}n_{R\mp}+\Omega P_{\pm}\left(  \mathbf{r}%
\right)  \right]  $ represents spin-conserved and spin-exchange interactions
with hot exciton reservoir where $P_{\pm}\left(  \mathbf{r}\right)  $ is the
spatially dependent pumping rate and $g_{1}$, $h_{1}$, $\Omega>0$ are
phenomenological coefficients to be determined experimentally. $V_{ext}\left(
\mathbf{r}\right)  $ represents the static disorder potential in semiconductor
microcavities, which is typically chosen as the same for both component
polaritons. $g_{2}n_{R\pm}$ and $h_{2}n_{R\mp}$ are related with the
condensation rate in that growth of condensate are stimulated by hot excitons
with same spin or cross spin, respectively \cite{PorrasPRB2002}. $\beta_{2}$
and $f_{2}$ are the same-spin and cross-spin nonradiative loss rates, respectively.

The equation \ref{Pola_1} of condensate is coupled to a rate equation
describing the time evolution of density $n_{R\sigma}$ of incoherent hot
exciton as:%
\begin{equation}
\partial_{t}n_{R\pm}=-\Gamma n_{R\pm}-\left[  g_{2}|\psi_{\pm}|^{2}+h_{2}%
|\psi_{\mp}|^{2}\right]  n_{R\pm}+P_{\pm}, \label{Exc_1}%
\end{equation}
where the reservoir relaxation rate $\Gamma$ is much faster than that of
condensate $\Gamma\gg\gamma_{C}$ where the Gaussian pump laser $P_{\pm}=W$ is
assumed nonpolarized (corresponding to linear or horizontal polarization)
providing a sufficient large occupation in momentum space of incoherent hot
exciton. The stimulated emission of the hot exciton reservoir into condensate
is taken into account by the term $\left[  g_{2}|\psi_{\pm}|^{2}+h_{2}%
|\psi_{\mp}|^{2}\right]  n_{R\pm}$. The spatial diffusion rate of reservoir
density has been neglected. In the following, we solve the coupled Eqs.
\ref{Pola_1} and \ref{Exc_1} numerically starting from a small random initial
condition. As we can see that, the time evolution of the system has been
obtained until a steady state is reached independent of the initial noise.

\begin{figure}
[t]%
\includegraphics[width=0.9\linewidth]{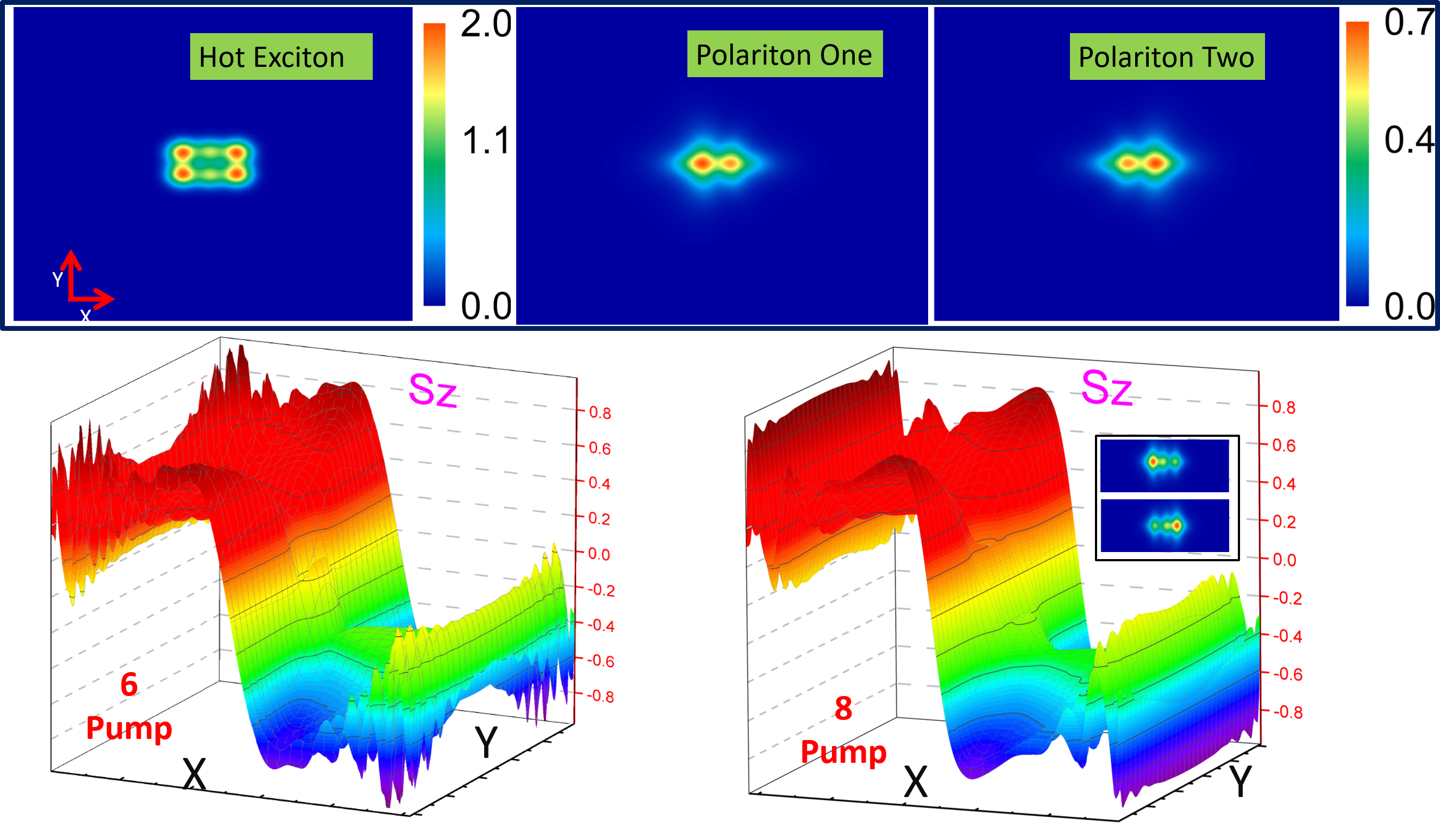}\caption{(Color online) The
spontaneously circular polarization of spinor condensate non-resonant pumped
by 6 and 8 linearly polarized laser, respectively. Top panel: density
distribution of hot exciton (left picture which has same profile for both
components) and spinor polariton (middle and right pictures for each
components) in real space for 6 pumping points. Bottom panel: distribution of
magnetic polarization along the Z axis for 6 pumping points (left picture) and
8 pumping points (right picture, where inset shows density distribution of two
component polariton). The size of profile is 24x24 and the other parameters used
in the simulations are the same as those in the Fig. \ref{Fig0}.}\label{Fig1}
\end{figure}

\section{Steady State.}

\subsection{Spatially homogeneous system}

Let us begin with some analytical consideration on spinor condensate. In the
homogeneous case, i.e., under a spatially homogeneous pumping and in the
absence of any external potential, Eqs. \ref{Pola_1} and \ref{Exc_1} admit
analytical stationary spinor configuration. Below the pumping threshold, the
condensate remains unpopulated, while the reservoir grows linearly with the
pump intensity as $n_{R\pm}=W/\Gamma$. At the threshold pump intensity
$W^{th}$, the stimulated emission rate exactly compensates the losses
$g_{2}n_{R\pm}+h_{2}n_{R\mp}=\gamma_{C}$ and condensate becomes populated
dynamically. We notice that threshold pump intensity becomes $W^{th}%
=\Gamma\gamma_{C}/\left(  g_{2}+h_{2}\right)  $. Above the threshold, the
reservoir density is homogeneous $n_{R\pm}=W/\left(  \Gamma+g_{2}|\psi_{\pm
}|^{2}+h_{2}|\psi_{\mp}|^{2}\right)  $, from this, we obtain
\begin{equation}
Z_{R}\sim-\frac{W\left(  g_{2}-h_{2}\right)  }{\Gamma^{2}+\Gamma\left(
g_{2}+h_{2}\right)  n_{c}}Z_{C}, \label{Exc_2}%
\end{equation}
here, we have defined reservoir polarization $Z_{R}=n_{R+}-n_{R-}$, condensate
polarization $Z_{C}=|\psi_{+}|^{2}-|\psi_{-}|^{2}$ and condensate total
density $n_{c}=|\psi_{+}|^{2}+|\psi_{-}|^{2}$. As long as $g_{2}\neq h_{2}$,
condensate polarization is directly proportional to the reservoir polarization.

From the Eqs. \ref{Pola_1}, we find that the condensate density is%
\begin{equation}
n_{c}\sim\frac{\left(  W-W^{th}\right)  }{\gamma_{C}}\cdot\frac{1}{1-\frac
{1}{2}\left(  \frac{W}{W^{th}}+\frac{\beta_{2}+f_{2}}{g_{2}+h_{2}}\frac
{\Gamma}{\gamma_{C}}\right)  }, \label{Pola_Den}%
\end{equation}
and condensate polarization satisfy%
\begin{equation}
M_{C}Z_{C}=0. \label{Pola_Mag}%
\end{equation}
where%
\[
M_{C}=\left(  4Wg_{2}h_{2}+\Gamma^{2}\left(  \beta_{2}-f_{2}\right)
-\frac{W\Gamma^{2}\gamma_{C}^{2}}{W^{th}\cdot W^{th}}\right)  .
\]
Except very stringent condition $M_{C}=0$, otherwise, magnetization of
condensate is always zero, i.e., $Z_{C}=0$\ seen from the Eq. \ref{Pola_Mag}.
If assuming cross-spin radiative and nonradiative loss rates is negligible,
magnetization condition $M_{C}=0$ leads to following condition for pump laser
power%
\begin{equation}
W=\frac{\gamma_{C}^{2}}{\beta_{2}\left(  W^{th}\right)  ^{2}}=\frac{g_{2}^{2}%
}{\beta_{2}\Gamma^{2}}, \label{Pola_4}%
\end{equation}
therefore, considering necessary condition $W>W^{th}$, we find following
conditon should be satisfied for spontaneous magnetization of condensate,
\[
\frac{g_{2}^{3}}{\beta_{2}\gamma_{C}\Gamma^{3}}>1.
\]

If assuming condensate wave function takes the form $\psi_{\pm}\left(
\mathbf{r}\right)  =\sum\psi_{\mathbf{k}_{\pm}\omega_{\pm}}e^{i\left(
\mathbf{k}_{\pm}\mathbf{\cdot r-}\omega_{\pm}t\right)  }\sim\psi
_{\mathbf{0}\pm}e^{i\left(  \mathbf{k}_{\pm}\mathbf{\cdot r-}\omega_{\pm
}t\right)  }$, we find spectrum as%
\begin{equation}
\omega_{\pm}=\frac{\hbar k_{\pm}^{2}}{2m}+\tilde{\Omega}_{\pm}W,
\label{Pola_5}%
\end{equation}
where%
\begin{align*}
\tilde{\Omega}_{\pm}  &  =\Omega+\frac{\left(  \beta_{1}+f_{1}\right)
n_{c}\pm\left(  \beta_{1}-f_{1}\right)  Z_{C}}{2W}\\
&  +\frac{2\left(  g_{1}+h_{1}\right)  \Gamma+G\cdot n_{c}\pm H\cdot Z_{C}%
}{2\left[  \Gamma^{2}+\Gamma\left(  g_{2}+h_{2}\right)  n_{c}+A\right]  },
\end{align*}
here, wave vector $\mathbf{k}_{\pm}$ and frequency $\omega_{\pm}$ remains so
far undetermined, and coefficience $\ G=g_{1}g_{2}+g_{1}h_{2}+g_{2}h_{1}%
+h_{1}h_{2}$, $H=\left(  g_{1}h_{2}+g_{2}h_{1}-g_{1}g_{2}-h_{1}h_{2}\right)
$. However, from Eq. \ref{Pola_5}, we find frequency difference between two
component is given by%
\begin{equation}
\omega_{+}-\omega_{-}=\frac{\hbar\left(  k_{+}^{2}-k_{-}^{2}\right)  }%
{2m}+\Delta\tilde{\Omega}, \label{Pola_6}%
\end{equation}
here,%
\begin{align*}
\Delta\tilde{\Omega}  &  \sim Z_{C}\left\{  \left(  \beta_{1}-f_{1}\right)
/W\right. \\
&  \left.  -\left(  g_{1}-h_{1}\right)  \left(  g_{2}-h_{2}\right)  /\left[
\Gamma^{2}+\Gamma\left(  g_{2}+h_{2}\right)  n_{c}+A\right]  \right\}  ,
\end{align*}
where $A$ is high order term of density and polarization $A=\left(
g_{2}+h_{2}\right)  ^{2}\left(  n_{c}^{2}+Z_{C}^{2}\right)  /4$ which can be
dominant term for the large density and polarization. Interestingly, we can
see that energy gap is polarization dependence. In particular, when $\beta
_{1}\simeq f_{1}$ or large enough laser power $W$, polarization dependence of
frequency difference disappears.

\subsection{Local density and spin approximation}

In the presence of an inhomogeneous laser pump $W\left(  \mathbf{r}\right)  $
(or multiple pump $W_{i}\left(  \mathbf{r}\right)  $), much richer phenomena
will be represented, such as, spin domain formation, emergent magnetic
monopole, generation of half vortex and so on. Under inhomogeneous laser pump,
we thus look for stationary spinor polariton wave function as following form%
\begin{equation}
\Psi=\left(
\begin{array}
[c]{c}%
\psi_{+}\\
\psi_{-}%
\end{array}
\right)  =\sqrt{\rho\left(  \mathbf{r}\right)  }\zeta(\mathbf{r})e^{-i\left(
\phi\left(  \mathbf{r}\right)  -\omega_{\pm}t\right)  }, \label{Pola_Spinor}%
\end{equation}
where $\rho\left(  \mathbf{r}\right)  $ and $\phi\left(  \mathbf{r}\right)  $
are the local density and phase of the condensate, and $\zeta(\mathbf{r})$ is
spinor function. We are going to assume that the local pump imposes a boundary
condition for the spinor function at each pumping spot $r_{p}$: $\lim
_{r\rightarrow r_{p}}\zeta(\mathbf{r})=\lambda$, $\lim_{r\rightarrow r_{p}%
}\mathbf{k}_{C}\left(  \mathbf{r}\right)  =0$, here, we have defined local
condensate density wave vector $\mathbf{k}_{C}\left(  \mathbf{r}\right)
=\nabla_{\mathbf{r}}\phi\left(  \mathbf{r}\right)  $. In the following, the
dimensionless form of the model can be obtained by using the scaling units of
time, energy, and length as: $T=1/\gamma_{C}$, $E=\hbar\gamma_{C}$,
$L=\sqrt{\hbar/m\gamma_{C}}$, respectively.

Inserting Eq. \ref{Pola_Spinor} into the Eqs. of motion \ref{Pola_1} and
\ref{Exc_1}, one obtains the following set of conditions for stationary
solution:%
\begin{align}
\omega_{\pm}  &  =-\frac{1}{2}\left(  \frac{\nabla^{2}\sqrt{\rho}}{\sqrt{\rho
}}+\frac{\nabla^{2}\zeta_{\pm}}{\zeta_{\pm}}+2\frac{\nabla\sqrt{\rho}%
\cdot\nabla\zeta_{\pm}}{\sqrt{\rho}\zeta_{\pm}}-k_{C}^{2}\right) \nonumber\\
&  +\frac{1}{\gamma_{C}}\left(  \beta_{1}|\zeta_{\pm}|^{2}\rho+f_{1}%
|\zeta_{\mp}|^{2}\rho+g_{1}n_{R\pm}+h_{1}n_{R\mp}\right)  +\frac{\Omega
W}{\gamma_{C}}, \label{Pola_R}%
\end{align}
and%
\begin{gather}
\frac{1}{2}\left(  g_{2}n_{R\pm}+h_{2}n_{R\mp}+\beta_{2}\left\vert \zeta_{\pm
}\right\vert ^{2}\rho+f_{2}|\zeta_{\mp}|^{2}\rho-\gamma_{C}\right) \nonumber\\
+\frac{1}{2}\nabla\cdot\mathbf{k}_{C}\left(  \mathbf{r}\right)  +\frac
{\nabla\sqrt{\rho}\cdot\mathbf{k}_{C}\left(  \mathbf{r}\right)  }{\sqrt{\rho}%
}+\frac{\mathbf{k}_{C}\left(  \mathbf{r}\right)  \cdot\nabla\zeta_{\pm}}%
{\zeta_{\pm}}=0, \label{Pola_I}%
\end{gather}
and%
\begin{equation}
\Gamma n_{R\pm}+\left(  g_{2}|\zeta_{\pm}(\mathbf{r})|^{2}+h_{2}|\zeta_{\mp
}(\mathbf{r})|^{2}\right)  \rho\left(  \mathbf{r}\right)  n_{R\pm
}=W(\mathbf{r}). \label{Exc_R}%
\end{equation}
Different from the single component condensate, now in Eq. \ref{Pola_R}, the
quantum pressure terms are not only originated from density $\nabla^{2}%
\sqrt{\rho}$ but also from the spinor $\nabla^{2}\zeta$ and even spin-density
coupling $\nabla\sqrt{\rho}\cdot\nabla\zeta$. Moreover, in Eq. \ref{Pola_I},
besides the current divergence term, we can see the more terms appeared which
is originated from coupling of superfluid current with density pressure
$\nabla\sqrt{\rho}\cdot\mathbf{k}_{C}\left(  \mathbf{r}\right)  $ or spin
pressure $\mathbf{k}_{C}\left(  \mathbf{r}\right)  \cdot\nabla\zeta$.

We can make local density approximation (LDA) and local spin approximation
(LSA) if the spatial variation of the laser pump $W\left(  \mathbf{r}\right)
$ is smooth enough. In such approximations, the quantum pressure term in Eq.
\ref{Pola_R} and \ref{Pola_I} can be neglected. Interestingly, similar to the
homogeneous case, the condensate density profile and polarization is still
given by the same Eq. \ref{Pola_Den} and Eq. \ref{Pola_Mag}, respectively,
except homogeneous laser pump $W$ is replaced with local value $W\left(
\mathbf{r}\right)  $ in there.

Under the Gaussian laser pump profile, we can look for cylindrically symmetric
stationary solutions. The condensate frequency $\omega_{\pm}$ is%
\begin{equation}
\omega_{\pm}=\frac{\tilde{\Omega}_{\pm}\cdot W}{\gamma_{C}}, \label{Pola_FR}%
\end{equation}
which is determined by the boundary condition that the local density wave
vector vanishes $\mathbf{k}_{C}\left(  \mathbf{r=r}_{p}\right)  =0$ at the
center of the each pumping spot. Here,%
\begin{align}
\tilde{\Omega}_{\pm}  &  =\Omega+\frac{\left(  \beta_{1}+f_{1}\right)  \rho
\pm\left(  \beta_{1}-f_{1}\right)  \rho S_{Z}}{2W}\nonumber\\
&  +\frac{2\left(  g_{1}+h_{1}\right)  \Gamma+\left[  G\cdot\rho\pm H\cdot\rho
S_{Z}\right]  }{2\left[  \Gamma^{2}+\Gamma\left(  g_{2}+h_{2}\right)
\rho+B\cdot\rho^{2}\right]  }, \label{Pola_FRCO}%
\end{align}
from here, we can find frequency difference between two component as%
\begin{equation}
\omega_{+}-\omega_{-}=\frac{\Delta\tilde{\Omega}\cdot W}{\gamma_{C}},
\label{Pola_FRDF}%
\end{equation}
here,
\begin{align*}
\Delta\tilde{\Omega}  &  =\tilde{\Omega}_{+}-\tilde{\Omega}_{-}\\
&  =\rho\left(  \mathbf{r}_{p}\right)  S_{Z}\left(  \mathbf{r}_{p}\right)
\left\{  \left(  \beta_{1}-f_{1}\right)  /W\right. \\
&  \left.  -\left(  g_{1}-h_{1}\right)  \left(  g_{2}-h_{2}\right)  /\left[
\Gamma^{2}+\Gamma\left(  g_{2}+h_{2}\right)  \rho+A\rho^{2}\right]  \right\}
,
\end{align*}
here, we have defined condensate polarization $S_{Z}\left(  \mathbf{r}%
_{p}\right)  =|\zeta_{+}\left(  \mathbf{r}_{p}\right)  |^{2}-|\zeta_{-}\left(
\mathbf{r}_{p}\right)  |^{2}$ and coefficient of density square term
$B=\left(  g_{2}+h_{2}\right)  ^{2}\left(  1+S_{Z}^{2}\right)  /4$, which has
maximal value $\left(  g_{2}+h_{2}\right)  ^{2}/2$ for the total polarization
$\pm1$. Interestingly, we can see that energy gap is polarization dependence.
In particular, when $\beta_{1}\simeq f_{1}$ or large enough laser power,
polarization dependence of frequency difference disappears.

Local density wave vector $\mathbf{k}_{C}\left(  \mathbf{r}\right)  $ of
condensate is reaching maximal value with the condensate density decreased and
spin polarized away from the pumping center. Polaritons condense at the laser
spot position has a large blueshifted energy due to their interactions with
uncondensed hot excitons, thus within a short time, these interaction energy
will lead to the motion of polariton initially localized at pumping point. In
particular, spontaneous polarization may happen because polarization may lower
the frequency obviously under the laser power is large enough as we can see
from Eq. \ref{Pola_FRCO}. Therefore, spin domain, spin current and topological
defect may be formed under such appropriate condition.

In the following, through extensive numerical simulations of the Eq.
\ref{Pola_1} coupled to the reservoir evolution Eq. \ref{Exc_1}, above
analytical results have been approved, such as, the dynamical formation of
spin domain, spin current and half vortex for a wide range of pump parameters
obviously available within state-of-the-art techniques.

\section{Numerical Results for Spontaneous Polarization.}

Eqs. of motion \ref{Pola_1} and \ref{Exc_1} can be solved numerically with the
initial condition $n_{R\sigma}(x,y,t)\approx0$, $\psi_{\sigma}(x,y,t)\approx
0$. The parameters of the pump are chosen according to the related experiments
\cite{NalitovPRL2015,DufferwielPRL2015,SalaPRX2015,HivetNphys2012} which study
the optical spin hall effect, tunable spin textures and half solitons. In our
calculations the following parameters are used typically for state-of-the-art
GaAs-based microcavities: the polariton mass is set to $m=$ $10^{-4}$
$m_{\text{e}}$ where $m_{\text{e}}$ is the free electron mass; the decay rates
are chosen as $\gamma_{C}=$ $0.152$ ps$^{-1}$ and $\Gamma$ $=$ $3.0\gamma_{C}%
$; thus, the scaling units of time, energy and length are $6.58$ ps, $0.1$
meV, and $1.\,\allowbreak54$ $\mu$m, respectively; the interaction strengths
are set to $\hbar\beta_{1}=$ $40$ $\mu$eV $\mu$m$^{2}$, $f_{1}=$
$-0.1\beta_{1}$, $g_{1}=$ $2\beta_{1}$, $h_{1}=$ $-0.2\beta_{1}$; the
condensation rate are set to $\hbar g_{2}=0.16$ meV $\mu$m$^{2}$, $\hbar
h_{2}=0.016$ meV $\mu$m$^{2}$, and condensation loss rate $-\hbar\beta
_{2}=0.16$ meV $\mu$m$^{2}$, $\hbar f_{2}=0.016$ meV $\mu$m$^{2}$. In our simulation,
the dimensionless scattering coefficient for each interaction term has been
tuned carefully in order to get the physical phenomena we want due to complicated nonlinear effects.
From an experimental point of view, the dimensionless interaction parameters must be adjusted to match pump intensity.
The pump intensity was chosen according to the experimentally measured blueshift of the
polariton condensate, and its profile is Gaussian shape as:%

\[
W(\mathbf{r})=\frac{w_{0}}{\pi w_{1}^{2}}%
%TCIMACRO{\dsum \nolimits_{i=1}^{n}}%
%BeginExpansion
{\displaystyle\sum\nolimits_{i=1}^{n}}
%EndExpansion
e^{\frac{-\left(  x-x_{i}\right)  ^{2}-\left(  y-y_{i}\right)  ^{2}}{w_{1}%
^{2}}},
\]
here, for a typical case, $w_{1}=1.0$, $\left\vert x_{i}\right\vert
=\left\vert y_{i}\right\vert =1.5$, and $w_{0}$ is tuned accordingly.

As expected, Eqs. of motion \ref{Pola_1} and \ref{Exc_1} tend to settle to a
steady state with a spontaneously circular polarization under increasing laser
power as shown in Fig. \ref{Fig0}(b). Threshold laser power for spontaneously
circular polarization is greater than that of starting condensation which can
be understood from our derived Eqs. \ref{Pola_Mag} and \ref{Pola_4}. The
coherent polarized polaritons ballistically fly away from the laser spot due
to their interactions converted into kinetic energy of coherent polariton. In
particular, the circular polarization rapidly saturates with increasing the
pumping power and may lead to an almost full polarization
\cite{BaumbergPRX2015,BaumbergPRL2016,AskitopoulosPRB2016}. Surprisingly, full
circular polarization state will change immediately back to the linear
polarization with further increasing the laser power (i.e., the density of
condensate exceeding a threshold value). Such phenomenon can be understood
from Eqs. \ref{Pola_R} and \ref{Pola_I} where various quantum pressure terms
will take important roles. Moreover, as shown in the Fig. \ref{Fig0}(c),
density profiles of incoherent hot exciton and polariton condensate represent
linear and circular polarization, respectively within spontaneously circular
polarization regime. As we can see that, while unpolarized hot excitons
experience a limited diffusion, polarized polaritons ballistically fly away
from the laser spot due to the conversion between interactions energy and
kinetic energy.

Fig. \ref{Fig1} show the density distribution of incoherent hot exciton and
coherent polariton condensate under six and eight unpolarized pumping laser
points. Interestingly, the neighbouring condensed polaritons are polarized
with opposite polarization as can be seen from $s_{z}$ distribution clearly.
Moreover, steady state with magnetic domain wall formation has been obtained
and characterized by vanishing total magnetization. Such phenomena is
fundamentally related with emergent effective magnetic field by the
inhomogeneous pump laser as we mentioned before. Furthermore, other
interesting magnetic textures may be formed from the evolution of Eqs. of
motion \ref{Pola_1} and \ref{Exc_1}. In the following, we will address the
question how to generate the density current, spin current, phase fluctuation
and slip via tuning pumping (or geometrical) source.

\begin{figure}
[t]\includegraphics[width=0.9\linewidth]{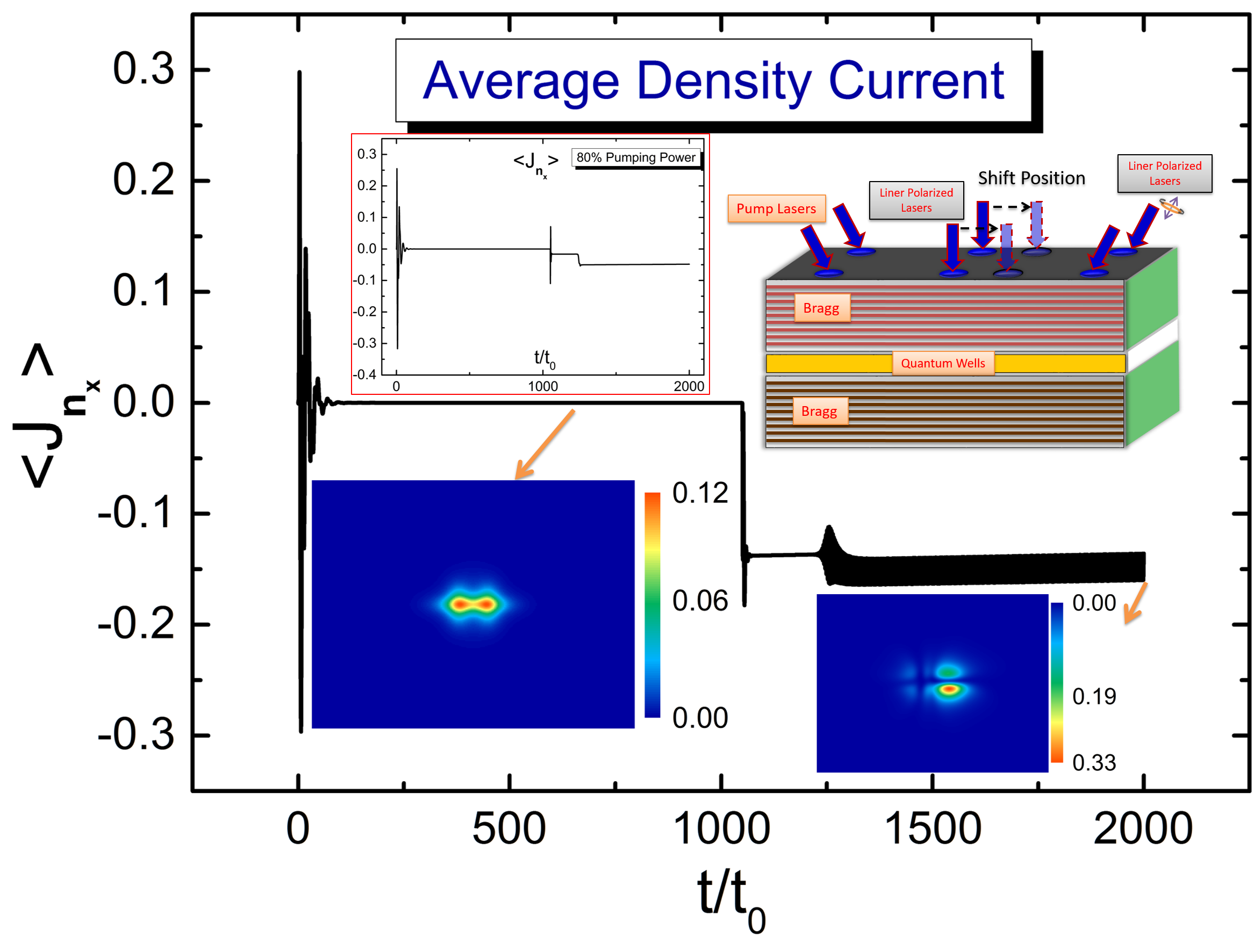}\caption{(Color online)
Normalized average density current $J_{n_{x}}$ of condensate which is
non-resonantly excited by 6 points laser with linear polarization. The insets
shows the total density profile before and after shifting position of two
middle lasers (see the schematic picture) along the x direction, and also
shows $J_{n_{x}}$ under decreasing the pumping power to 80\%.
The size of profile is 24x24 and the other parameters used
in the simulations are the same as those in the Fig. \ref{Fig0}.}\label{Fig2}
\end{figure}

\section{Density Current, Spin Current, Phase Slip.}

Physically, condensed fluid is a long-range cooperative phenomenon
characterized by long-range correlation and coherent ordering of the momenta
of particle. The various correlation function may imply net surface currents
and orbital angular momentum appearing in this system. Therefore, It is
important to study the density and spin current, and furthermore, study how to
generate and control them. In the following, we will address these questions
by suddenly shifting pumping laser position by a distance. Interestingly, we
find that a steady current can be generated apparently. In particular, if
shifted the pumping laser is linear or circular polarized, we observe large
phase fluctuation where ring-shape phase jump shows the behavior of splitting
and joining together. The above-mentioned behaviors may be understood from
emergent effective gauge field caused by externally pumped incoherent reservoirs.

\subsection{Density current}

First, we numerically simulate time evolution of Eqs. of motion \ref{Pola_1}
and \ref{Exc_1} under pumped by six linearly polarized laser and then,
suddenly shifting two middle laser's position. The results are shown in the
Fig. \ref{Fig2} for the average density current, which are normalized by the
total density of condensate. As is shown in the Fig. \ref{Fig2}, a
large-amplitude oscillation appears within a short time when switching on the
pumping lasers. With time evolution, oscillation decays very quickly and
disappear at 60 unit of time. The appearance of such oscillation can be
understood from the large overlap of incoherent hot exciton and coherent
polariton which leads to the large repulsive force in the beginning. Then,
with coherent polariton's diffusion under such repulsive force, condensate
stay in a steady state with zero averaged current, which means a balanced
configuration in momentum space of condensate.

Second, we want to generate steady current without decay by breaking above
balanced configuration. Therefore, we suddenly shift two middle laser's
position at time 1050 (referring to the schematic picture in the inset of Fig.
\ref{Fig2}). Interestingly, a persistent current with small oscillation can be
observed clearly and it's amplitude is centred at -0.15. The appearance of
such persistent current can be understood from breaking balanced-momentum
configuration due to changing interaction energy between different part of
condensate. Moreover, accompanied fast small-amplitude oscillation can be
understood as surface oscillation modes which are moved back and forth due to
confinement by the pumping laser. Furthermore, such oscillation can be
suppressed by lowering the pumping power completely as is shown in the inset
of Fig. \ref{Fig2}, where pumping power drops up to 80 percent of previous
case. However, we can not generate persistent spin current by using above
method. Therefore, we will address this issue in the following section.

\begin{figure}
[t]\includegraphics[width=0.9\linewidth]{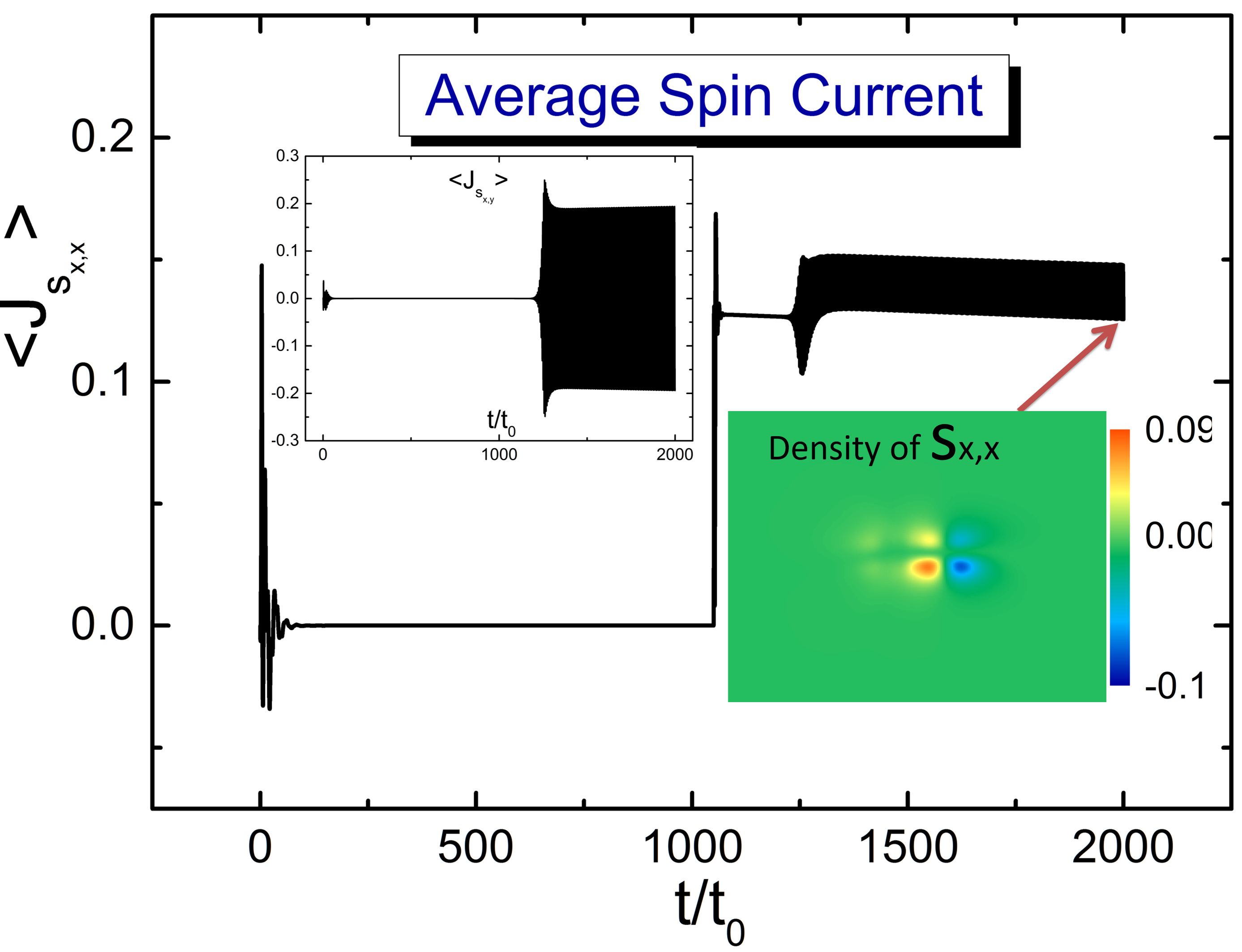}\caption{(Color online)
Normalized average spin current $\left\langle J_{s_{x,x}}\right\rangle $ of
condensate which is non-resonantly excited by 6 points laser with linear
polarization. The insets shows density profile of the spin current
$J_{s_{x,x}}$ at the final stage after shifting position of two middle lasers
along the x direction, and that for normalized average spin current along the
y direction $\left\langle J_{s_{x,y}}\right\rangle $. The size of profile is 24x24 and the other parameters used
in the simulations are the same as those in the Fig. \ref{Fig0}.}\label{Fig3}
\end{figure}

\begin{figure}
[t]\includegraphics[width=0.9\linewidth]{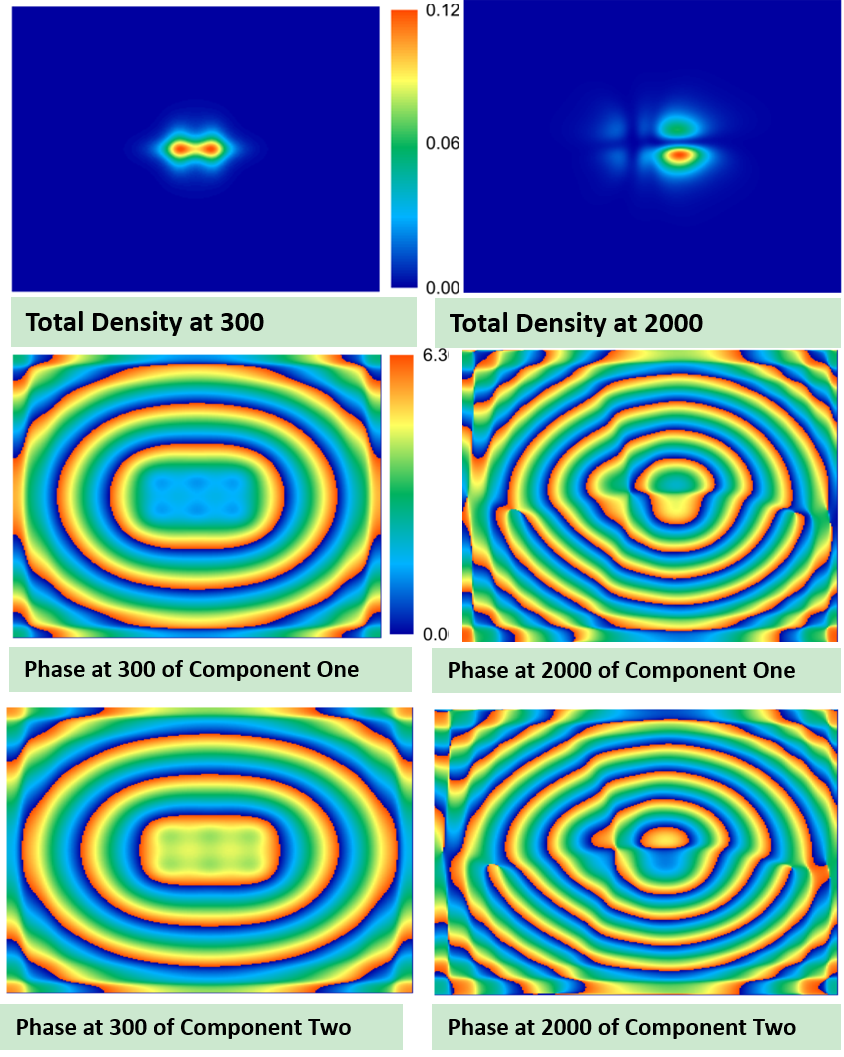}\caption{(Color online)
Total density and each phase profile of condensed polariton excited by 6
points laser with linear polarization. The left column and right column
correspond to the spatial distributions before and after shifting position of
two middle lasers along the x direction, respectively. The size of profile is 24x24 and the other parameters used
in the simulations are the same as those in the Fig. \ref{Fig0}.}\label{Fig4}
\end{figure}

\begin{figure}
[t]\includegraphics[width=0.9\linewidth]{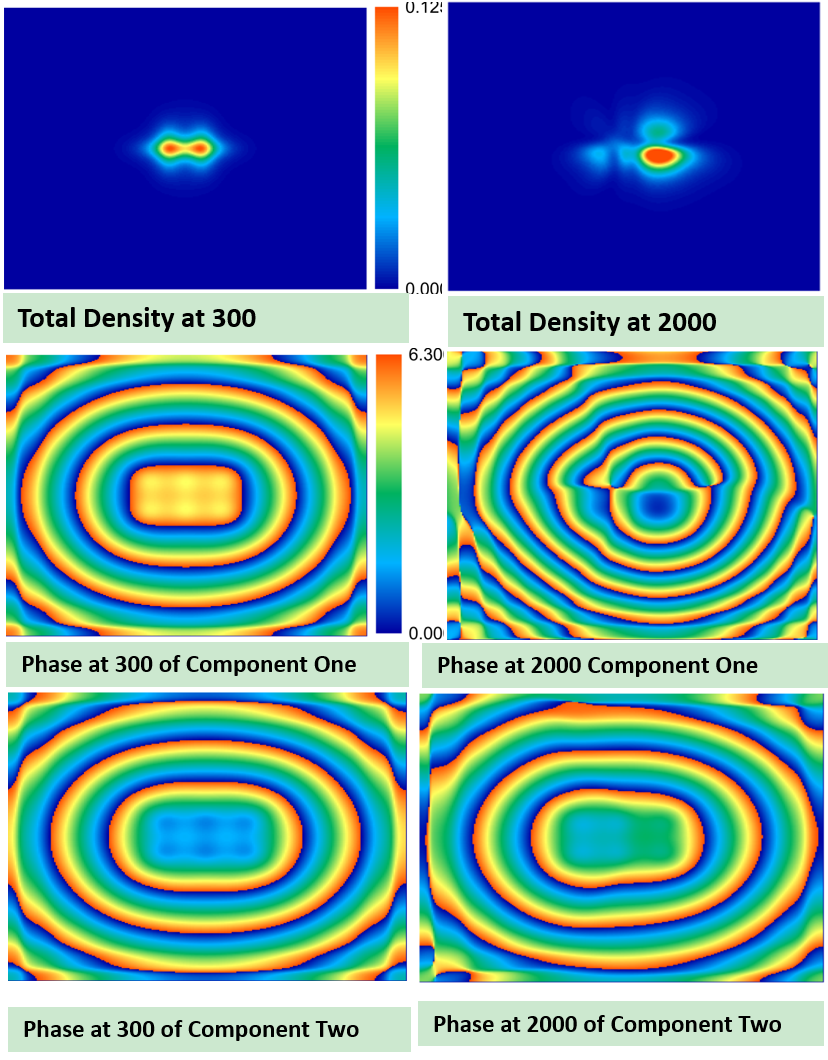}\caption{(Color online)
Total density and each phase profile of condensed polariton excited by 6
points laser with circular polarization. The left column and right column
correspond to the spatial distributions before and after shifting position of
two middle lasers along the x direction, respectively. The size of profile is 24x24 and the other parameters used
in the simulations are the same as those in the Fig. \ref{Fig0}.}\label{Fig5}
\end{figure}

\subsection{Spin current}

Polariton condensates are excellent candidates for designing novel
spin-based\ devices at room temperature due to their many features, such as
strong optical nonlinear response, spin polarization properties, and fast spin
dynamics. Therefore, in the following, we will show how to generate spin
transportation of coherent polariton. In particular, we observed persistently
long-range spin transport without dissipation. We will show our results
obtained by numerically simulate time evolution of Eqs. of motion \ref{Pola_1}
and \ref{Exc_1} in the following.

First, we obtained time evolution of average spin current $\left\langle
J_{s_{x,x}}\right\rangle $ as shown in the Fig. \ref{Fig3}, where polariton
condensate is non-resonantly excited by 6 pumping lasers with linear
polarization. In the early stage, a large-amplitude oscillation appears within
a short time when switching on the pumping lasers. With time evolution,
oscillation decays very quickly and disappear at 60 unit of time. Above
phenomena are very similar to those of average density current shown in
Figures \ref{Fig2}. However, it is interesting to point out that such
large-amplitude oscillation has very asymmetric behavior in contrary to
symmetric behavior in average density current. Such asymmetric phenomenon may
be understood from the symmetry breaking by effective magnetic field
stimulated by the pumping lasers. Importantly, the remaining question is how
to generate steady spin current without dissipation. Therefore, we try to deal
with such question by manipulating pumping laser.

Interestingly, persistent spin current is quickly developed at time 1050 and
it's amplitude is centred at 0.15. Moreover, the fast small-amplitude
oscillation still appear which may be understood as stimulating surface
oscillation mode by breaking symmetry on the spatial distribution of pumping
lasers. Next, we compare the spin current along the different directions.
Interestingly, average spin current $\left\langle J_{s_{x,y}}\right\rangle $
along the y direction has dramatically different behavior as shown in the
inset of Fig. \ref{Fig3}. As we can see that $\left\langle J_{s_{x,y}%
}\right\rangle $ represents very symmetric oscillation centred at zero value.
Such different behavior between $\left\langle J_{s_{x,x}}\right\rangle $ and
$\left\langle J_{s_{x,y}}\right\rangle $ is due to shift lasers' position
along the x direction instead of y direction. Therefore, we can conclude that
net spin current may be induced by breaking symmetric distribution of pumping
lasers along preferred direction. It must be pointed out that local spin
current $S_{x,x}$ may be positive or negative value as indicated in the insets
of Fig. \ref{Fig3}. Such nondissipative spin current is induced by the
effective magnetic field with density- or current-dependence function.
Importantly, manipulation of such effective magnetic field may be utilized to
generate various polarization textures as well as spin-polarized vortices.
Now, the question is how to generate stable topological defects in our studied system.

\subsection{Phase Slip}

As is well known, condensed polariton provides a very promising platform to
generate and control spin current and various spin textures through
manipulating effective gauge fields (like Dresselhaus and Rashba fields). In
particular, there are many kinds of quantum phases in spinor quantum fluids
can be accessible experimentally in this platform. For example, there may
generate fascinating topological defects by manipulating pumping lasers
\cite{LagoudakisNphys2008,RoumposNphys2011,SanvittoNphot2011,BoulierPRL2016}.

Physically, in order to generate topological defects, large phase fluctuations
must be occurred by reducing the coherence length and amplitude of the order
parameter (polariton condensate). Therefore, let us first study how to
generate large phase fluctuations. In order to generate that, we suddenly
moved the position of the pumping lasers in the middle site, and then see how
the phase fluctuations are formed dynamically.

Figures \ref{Fig4} shows the total density and each phase profile of condensed
polariton before and after moving the lasers in the middle site, where each
component of condensed polariton is illuminated with the same laser power.
Interestingly, while condensed polaritons are concentrated on the right part,
large phase disturbance has been generated for each component. In particular,
in low density region, there are large phase fluctuations where ring-shape
phase jump shows the behavior of splitting and joining together. Physically,
due to energy advantages, topological defects are initially formed in
low-density regions, then, due to the dissipation of energy, these topological
defects gradually moved to the high-density area and eventually reached a
stable state. Therefore, we can expect that it is very promising to produce
stable topological defects (such as quantum vortices) in such system.

Furthermore, we want to control which component of the condensed polaritons
will generate large phase fluctuations. Therefore, each component of condensed
polariton is illuminated with the different laser power and then see how the
phase fluctuations are formed dynamically. Figures \ref{Fig5} shows the total
density and each phase profile of condensed polariton before and after moving
the lasers in the middle site. Interestingly, the phases of the two components
have very different shape distributions. Here, large phase disturbance has
been generated for component one which was illuminated with the laser power,
however, there is not much change in the phase of the second part. Moreover,
in component one, large phase fluctuations are closer to high-density areas
where ring-shape phase jump shows the behavior of splitting and joining together.

It must be admitted that stable topological defects are not created as they
require reconfiguring a large number of spins and density at a large energy
cost. Generally, what kinds of stable topological defects are developed
depending on the dynamics of gauge potential together with vector field, such
as Maxwell-Chern-Simons-vector Higgs model for the the superconductivity of
Sr$_{2}$RuO$_{4}$ \cite{MackenzieRMP2003}. In our studied non-equilibrium
exciton-polaritons liquid, spin- and density-dependent effective gauge fields
play important roles on the phase fluctuations and make effective gauge fields
more controllable comparing with conventional solid state system and ultracold
atoms. Finally, we remark that the physics described in our study may be
generally applicable to the recovery of complex order parameters in other
systems. Photoinduced phase fluctuations may be crucial to understanding the
mechanism of photoinduced superconductivity in the striped cuprates. These
phenomena can be conveniently probed by real-space spectroscopy, and phase
imaging \cite{LagoudakisScience2009}.

\section{Conclusions.}

In conclusion, we have demonstrated a practical way to control spin
polarization, generate density and spin current, and induce large phase
fluctuations in an exciton-polariton condensate. For the polariton lifetime,
The above-mentioned behaviors can be readily excited in photoluminescence
experiments and detectable by the time-resolved micro-photoluminescence
spectroscopy \cite{LagoudakisNphys2008,Amo2009} or spin noise
spectroscopy~\cite{ZapasskiiAOP2013,OestreichPSS2014}. Our results are of
particular significance for creating these excitations in experiments and for
exploring novel phenomena associated with them. This noticeably spin
amplification and spin transport could offer a promising way to optimize spin
signals in future devices with using polariton condensates.

\vspace{2mm}

\textbf{Acknowledgments }We are grateful to N. Berloff for discussions. The
financial support from the early development program of NanChang University
and Skoltech-MIT Next Generation Program is gratefully acknowledged.

\vspace{2mm}

\textbf{Compliance with ethical standards}

\vspace{2mm}

\textbf{Conflict statement} We declare we have no conflict of interests.

%\end{widetext}

\end{document}